\documentclass[twoside,fleqn,reqno]{article}

\usepackage[centertags,leqno,sumlimits,intlimits,namelimits]{amsmath}
\usepackage{amsfonts,amsbsy,amsxtra}
\usepackage{amssymb,latexsym}
\usepackage{graphicx}
\usepackage[latin1]{inputenc}
\usepackage[headings]{espcrc2} % with running headings

\newcommand{\antibar}[1]{%
  \mkern 3mu \overline{\mkern -3mu #1 \mkern -0.5mu} \mkern 0.5mu%
}
\newcommand{\PPh}{\ensuremath{\gamma}}
\newcommand{\PvPh}{\ensuremath{\gamma^*}}
\newcommand{\PPom}{\ensuremath{\mathbb{P}}}
\renewcommand{\Pr}{\ensuremath{\rho}}
\newcommand{\Prz}{\ensuremath{\rho^0}}
\newcommand{\Po}{\ensuremath{\omega}}
\newcommand{\Ppi}{\ensuremath{\pi}}
\newcommand{\Ppip}{\ensuremath{\pi^+}}
\newcommand{\Ppim}{\ensuremath{\pi^-}}

\newcommand{\Pep}{\ensuremath{e^+}}
\newcommand{\Pem}{\ensuremath{e^-}}
\newcommand{\Pepm}{\ensuremath{e^\pm}}
\newcommand{\Pqq}{\ensuremath{q}}
\newcommand{\Pqqbar}{\ensuremath{\antibar{q}}}
\newcommand{\Pp}{\ensuremath{p}}
\newcommand{\Pn}{\ensuremath{n}}

\newcommand{\Pd}{\ensuremath{d}}
\newcommand{\PA}{\ensuremath{A}}
\newcommand{\PAu}{\ensuremath{\text{Au}}}
\newcommand{\PAuex}{\ensuremath{\text{Au}^\ast}}
\newcommand{\sqrtsnn}{\ensuremath{\sqrt{s_{_{\!N\!N}}}}}
\newcommand{\mevc}{~\ensuremath{\text{MeV}\! / \!c}}

\newcommand{\gevcsq}{~\ensuremath{(\text{GeV}\! / \!c})^2}
\newcommand{\mevcc}{~\ensuremath{\text{MeV}\! / \!c^2}}

\newcommand{\lrbrk}[1]{{\left({#1}\right)}}
\newcommand{\lrBrk}[1]{{\left[{#1}\right]}}
\newcommand{\lrabs}[1]{{\left|{#1}\right|}}
\newcommand{\D}[2][]{\operatorname{d^{#1} \mathnormal{#2}}}
\renewcommand{\Re}{\mathfrak{Re}}

\newcommand{\pT}{\ensuremath{p_T}}
\newcommand{\figref}[1]{{Fig.~\ref{fig:#1}}}
\newcommand{\Figref}[1]{{Figure~\ref{fig:#1}}}
\newcommand{\equref}[1]{{Eq.~\eqref{eq:#1}}}
\newcommand{\Equref}[1]{{Equation~\eqref{eq:#1}}}

\newcommand{\Tabref}[1]{{Table~\ref{tab:#1}}}
\newcommand{\secref}[1]{{section~\ref{sec:#1}}}

\newcommand{\subsecref}[1]{{subsection~\ref{subsec:#1}}}

\newcommand{\others}{\textit{et al.}}
\newcommand{\ibid}{\textit{ibid.}}
\newcommand{\measresult}[4]{\ensuremath{#1 \pm #2_\text{stat.} \pm #3_\text{syst.}~\text{#4}}}

\title{Photoproduction in Ultra-Peripheral Heavy Ion Collisions at
  STAR}

\author{%
  B. Grube
  \address{Pusan National University, Department of Physics,
    30 Jangjeon-Dong Geumjeong-Gu, \\
    Busan 609-735, Republic of Korea}
  \address{Brookhaven National Laboratory, Physics Dept., Bldg. 510A,
    Upton, NY 11973, U.S.A.}
  for the STAR Collaboration
}

\runauthor{B. Grube}

\begin{document}

\begin{abstract}
  We present recent STAR results on photoproduction in
  ultra-peripheral relativistic heavy ion collisions. In these
  collisions the impact parameter of the beam particles is larger than
  the sum of their nuclear radii, so that they interact via their
  long-range Coulomb fields. STAR has measured the production of
  $\Prz(770)$ mesons in exclusive reactions as well as in processes
  with mutual nuclear excitation of the beam particles. We present
  results for the \Prz~production cross section in \PAu-\PAu\
  collisions at $\sqrtsnn = 200$~GeV for coherent as well as
  incoherent coupling. The dependence of the cross section on the
  \Prz~rapidity is compared to theoretical models. We also studied the
  ratio of coherent~\Prz\ to direct \Ppip\Ppim~production as well as
  the \Prz~helicity matrix elements and we observe interference
  effects in the \Prz~production. In addition STAR has measured the
  production of \Prz~mesons in \Pd-\PAu\ collisions at $\sqrtsnn =
  200$~GeV and that of \Pep\Pem-pairs in \PAu-\PAu\ at $\sqrtsnn =
  200$~GeV. We also see an enhancement around 1510\mevcc\ in
  \Ppip\Ppim\Ppip\Ppim\ final states in \PAu-\PAu\ collisions at
  $\sqrtsnn = 200$~GeV.
  \vspace{1pc}
\end{abstract}

\maketitle

\section{Introduction}
\label{sec:introduction}

Heavy ions are surrounded by a cloud of virtual photons. The
electromagnetic field of a relativistic nucleus can be approximated by
a flux of quasi-real virtual photons using the Weizs\"acker-Williams
approach~\cite{weizsacker_williams}. Since the number of photons grows
with the square of the nuclear charge, fast moving heavy ions can
generate large photon fluxes. Beams of heavy ions can thus be used as
photon sources or targets. Due to the long range of the
electromagnetic interactions, they can be separated from the hadronic
interactions by requiring impact parameters~$b$ larger than the sum of
the nuclear radii~$R_\PA$ of the beam particles. These so-called
Ultra-Peripheral heavy ion Collisions~(UPCs) allow to study
photonuclear effects as well as photon-photon
interactions~\cite{baur_krauss_bertulani_hencken}.

\begin{figure}[htb]
  \includegraphics[scale=0.443]{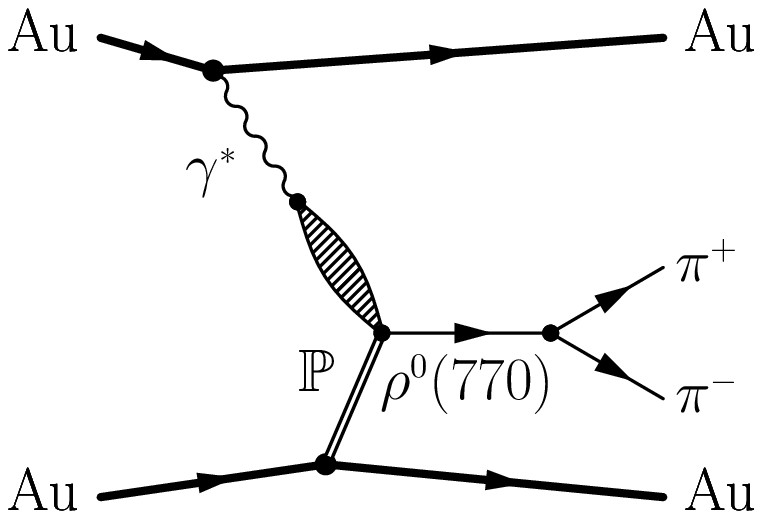}\hspace{0.5ex}
  \includegraphics[scale=0.443]{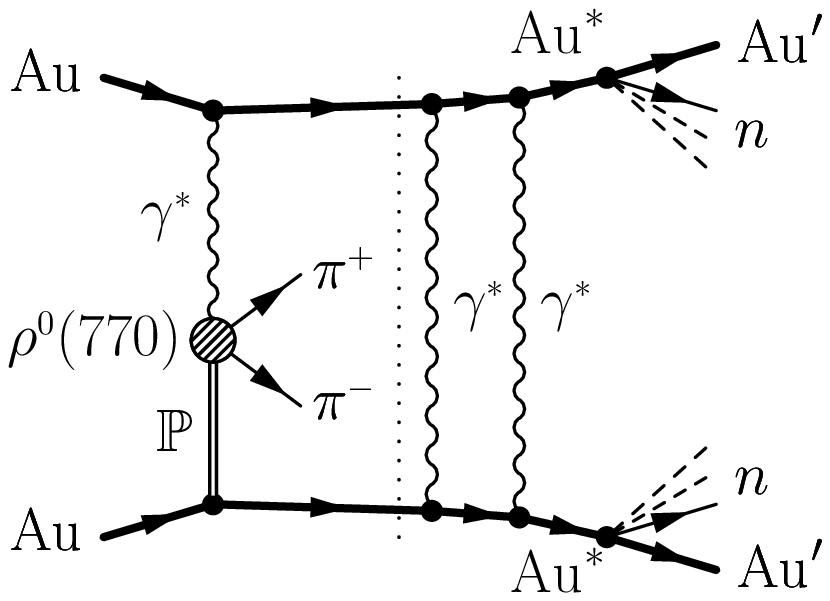}\\
  \centering a) \hspace{0.5\linewidth} b)
  \vspace{-3ex}
  \caption{(a)~Exclusive photonuclear \Prz~production in \PAu-\PAu\
    ultra-peripheral collisions. (b)~Photonuclear \Prz~production
    accompanied by mutual Coulomb excitation of the ions. The
    \Prz~creation in \PvPh\PPom~fusion is adumbrated in form of a
    hatched circle.}
  \label{fig:rho_prod}
\end{figure}

A typical photonuclear reaction is the production of vector mesons. In
this process the virtual photon, that is emitted by the ``spectator''
nucleus, fluctuates into a virtual \Pqq\Pqqbar-pair, which scatters
elastically off the ``target'' nucleus, thus producing a real vector
meson. The scattering can be described in terms of soft Pomeron
exchange. \Figref{rho_prod}a illustrates this reaction taking as an
example the most abundant process, the photonuclear production of
$\Prz(770)$ mesons.

The cross section for vector meson production depends on the way the
virtual \Pqq\Pqqbar-pair couples to the target nucleus. This is mainly
determined by the transverse momentum~\pT\ of the produced meson. For
small transverse momenta of the order of $\pT \lesssim 2 \hbar /
R_\PA$ the \Pqq\Pqqbar-pair couples coherently to the entire nucleus.
This leads to large cross sections which depend on the nuclear form
factor $F(t)$ and which scale roughly with the atomic number~$A$
squared. For larger transverse momenta the \Pqq\Pqqbar-pair couples to
the individual nucleons in the target nucleus. This ``incoherent''
scattering has a smaller cross section that scales approximately
with~$A$ modulo corrections for nuclear absorption of the meson.

Due to the intense photon flux, it is possible for the vector meson
production to be accompanied by Coulomb excitation of the beam
particles. The excited ions mostly decay via the emission of neutrons
which can be measured by the Zero Degree
Calorimeters~(ZDCs)~\cite{star_zdc} of the STAR~setup. To lowest order
events with mutual nuclear breakup are described by three-photon
exchange (see \figref{rho_prod}b): One photon to produce the vector
meson and two photons to excite the two nuclei~\cite{baltz_baur}. All
three photon exchanges are in good approximation independent, so that
the cross section for the production of a vector meson~$V$ can be
factorized:
\begin{multline}
  \sigma(\PAu\PAu \to \PAuex\PAuex\, V) = \\
    \int\!\D[2]{b} \lrBrk{1 - P_\text{had}(b)}\, P_V(b)\, P_{X\Pn, 1}(b)\, P_{X\Pn, 2}(b)
  \label{eq:sigma_mutual_excitation}
\end{multline}
where $P_\text{had}$~is the probability for hadronic interaction,
$P_V$~the probability to produce a vector meson~$V$, and $P_{Xn, i}$
the probability that nucleus~$i$ emits $X$~neutrons. Compared to
exclusive vector meson production, reactions with mutual Coulomb
excitation have smaller median impact parameters.

In this paper we present results from the STAR experiment at the
Relativistic Heavy Ion Collider~(RHIC). Charged tracks were
reconstructed using a large cylindrical Time Projection
Chamber~(TPC)~\cite{star_tpc} with 2~m radius and a length of 4.2~m,
that was operated in a 0.5~T solenoidal magnetic field. For tracks
with pseudorapidity $|\eta| < 1.2$ and transverse momentum $\pT >
100\mevc$ the tracking efficiency is better than 85~\%. For triggering
two detector systems were used: the Central Trigger
Barrel~(CTB)~\cite{star_ctb}, which is an array of 240~scintillator
slats surrounding the TPC that allows to trigger on charged particle
multiplicities in the event, and the two zero degree calorimeters,
which are located at $\pm18$~m from the interaction point and measure
neutrons originating from nuclear breakup.

\mathversion{bold}
\section{Photonuclear \Prz~Production}
\label{sec:rho_production}
\mathversion{normal}

STAR uses two kinds of triggers to select UPC events. The CTB-based
``topology'' trigger requires a low total charged particle
multiplicity and divides the CTB detector into four azimuthal
quadrants. Events are recorded when the left and right quadrants have
coincident hits, selecting roughly back-to-back pion pairs. The top
and bottom quadrants serve as a cosmic ray veto. The ``minimum bias''
trigger selects events where both nuclei dissociated by requiring
coincident energy deposits in both ZDCs.

In the offline analysis the events were required to have two tracks of
opposite charge that form a common (primary) vertex. In addition we
demanded a low overall track multiplicity in order to suppress
backgrounds from beam-gas interactions, peripheral hadronic
interactions, and pile-up events. The primary vertex was confined to a
region close to the interaction diamond which reduces contaminations
from pile-up events, beam-gas interactions, and cosmic rays. The
latter background component was decreased further by either excluding
events around $y_\Pr = 0$ in the topology data sample or by requiring
neutron signals in the ZDCs as in the minimum bias trigger. The
contamination from peripheral hadronic interactions was in addition
suppressed by demanding the track pair to have low net transverse
momentum. The backgrounds from two-photon \Pep\Pem\ (cf.
\secref{ee_production}) and photonuclear \Po~production are
negligible.

The \Ppip\Ppim\ invariant mass distributions are fit with the function
\begin{multline}
  \frac{\D{\sigma}}{\D{M_{\Ppi\Ppi}}} = 
  \lrabs{A\, \frac{\sqrt{M_{\Ppi\Ppi} M_\Pr \Gamma}} {M_{\Ppi\Ppi}^2 - M_\Pr^2 + i M_\Pr \Gamma} + B}^2 + f_\text{BG} \\
  \text{with} \quad \Gamma(M_{\Ppi\Ppi}) \equiv \Gamma_\Pr\, \frac{M_\Pr}{M_{\Ppi\Ppi}}
  \lrBrk{\frac{M_{\Ppi\Ppi}^2 - 4 m_\Ppi^2}{M_\Pr^2 - 4 m_\Ppi^2}}^\frac{3}{2}
  \label{eq:rho_mass}
\end{multline}

\Equref{rho_mass} consists of a relativistic Breit-Wigner with
amplitude~$A$, that describes the resonant \Ppip\Ppim\ production, and
a mass-independent amplitude~$B$ for the non-resonant \Ppip\Ppim\
production, as well as a S\"oding interference
term~\cite{soeding_term} of the two amplitudes. The second order
polynomial, $f_\text{BG}$, describes the combinatorial background
which was estimated from like sign pion pairs.

\mathversion{bold}
\subsection{Coherent \Prz~Production in \PAu-\PAu\ Collisions}
\label{subsec:rho_au_au_coh}
\mathversion{normal}

\begin{figure}[htb]
  \begin{center}
    \includegraphics[width=0.8\linewidth]{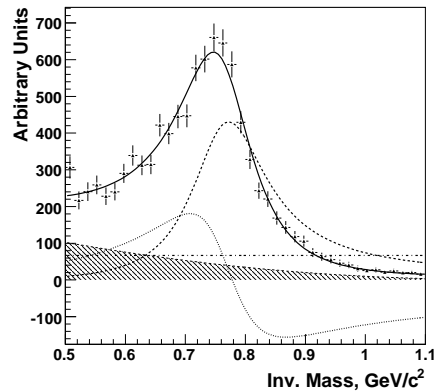}
  \end{center}
  \vspace{-5ex}
  \caption{\Ppip\Ppim\ invariant mass distribution for coherently
    produced~\Prz\ from the minimum bias data set~\cite{run2_rho}. The
    solid line is the fit result using \equref{rho_mass}, which has
    four components: The relativistic Breit-Wigner of the resonant
    \Ppip\Ppim\ production (dashed line), the mass-independent term
    for the non-resonant \Ppip\Ppim\ production (dash-dotted line),
    the interference of the two (dotted line), and the combinatorial
    background (hatched area).}
  \label{fig:run2_rho_mass_minbias}
\end{figure}

Coherent photonuclear \Prz~production is strongly enhanced at low
$\pT^\Pr$. It can be separated from the incoherent production by
requiring \Prz~transverse momenta smaller than 150\mevc.
\Figref{run2_rho_mass_minbias} shows the \Ppip\Ppim\ invariant mass
distribution of the minimum bias data set taken during the year 2002
run in \PAu-\PAu\ collisions at $\sqrtsnn = 200$~GeV~\cite{run2_rho}.
The peak contains $3\,075 \pm 128_\text{stat.}$~\Prz\ candidates.

Since in \equref{rho_mass} the amplitudes $A$ and $B$ are free fit
parameters, it is possible to extract the ratio $|B / A|$ of
non-resonant to resonant \Ppip\Ppim\ production. In \PAu-\PAu\
collisions at $\sqrtsnn = 200$~GeV the ratio was measured using the
minimum bias data to $|B / A| =
\measresult{0.89}{0.08}{0.09}{GeV$^{-1/2}$}$~\cite{run2_rho}. As
expected the ratio shows no dependence on the \Prz~rapidity nor on the
\Prz~decay angles. The value is compatible with an earlier result of
$|B / A| = \measresult{0.81}{0.08}{0.20}{GeV$^{-1/2}$}$ from the year
2000 run at $\sqrtsnn = 130$~GeV~\cite{run1_rho}. Both values are also
in agreement with the ZEUS measurements~\cite{zeus_rho}.

The luminosity of the minimum bias sample was measured based on the
known total hadronic production cross section of 7.2~barn. With the
value of $L_\text{minbias} = 461$~mb$^{-1}$ the cross section for
coherent \Prz~production accompanied by mutual nuclear excitation was
estimated to be \measresult{14.5}{0.7}{1.9}{mb$^{-1}$} within the
experimental acceptance of $|y_\Pr| < 1$~\cite{run2_rho}.

\begin{figure}[htb]
  \begin{center}
    \includegraphics[width=0.8\linewidth]{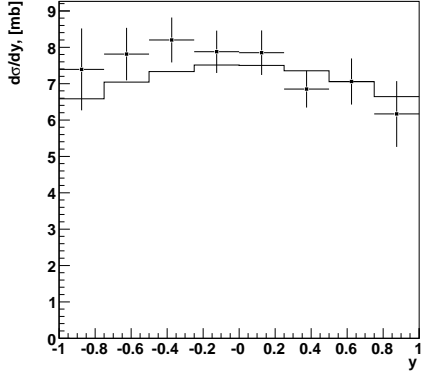}
  \end{center}
  \vspace{-5ex}
  \caption{Differential cross section $\D{\sigma}\! /\! \D{y_\Pr}$ for
    coherent \Prz~production accompanied by mutual Coulomb excitation
    as a function of \Prz~rapidity extracted from the minimum bias
    data~\cite{run2_rho}. The solid line shows the KN~model
    calculation~\cite{KN}.}
  \label{fig:run2_rho_y_dep_coh}
\end{figure}

There are at least three models for \Prz~production in
ultra-peripheral collisions: The model of Klein and
Nystrand~(KN)~\cite{KN} uses the Vector Dominance Model~(VDM) to
describe the virtual photon and a classical mechanical approach for
the scattering on the target nucleus, based on results from $\PPh \Pp
\to \Prz\Pp$ experiments. The Frankfurt, Strikman, and Zhalov~(FSZ)
model~\cite{FSZ} employs a generalized VDM for the virtual photon and
a QCD Gribov-Glauber approach for the scattering.  The model of
Gon\c{c}alves and Machado~(GM)~\cite{GM} is based on a QCD color
dipole approach that takes into account nuclear effects and parton
saturation phenomena. Only the KN~model gives predictions for the
rapidity dependence of the coherent \Prz~production cross section
accompanied by mutual Coulomb excitation, which are compared to the
data in \figref{run2_rho_y_dep_coh}.

\begin{figure}[htb]
  \begin{center}
    \includegraphics[width=0.8\linewidth]{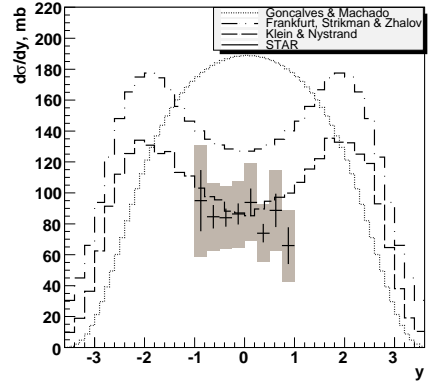}
  \end{center}
  \vspace{-5ex}
  \caption{Comparison of the measured cross section for coherent
    \Prz~production with theoretical predictions~\cite{run2_rho}. The
    vertical line at each point shows the statistical error. The
    shaded area displays the sum of statistical and systematical
    errors. The dashed line represents the KN~\cite{KN}, the
    dash-dotted line the FSZ~\cite{FSZ}, and the dotted one the
    GM~model~\cite{GM}.}
  \label{fig:run2_rho_y_dep}
\end{figure}

The total coherent \Prz~production cross section was estimated using
the $13\,054 \pm 124_\text{stat.}$~\Prz~candidates in the data set
taken with the topology trigger~\cite{run2_rho}, which does not apply
any restrictions on the ZDC amplitudes. Although this data sample
could not be used for an absolute cross section measurement, because
the trigger efficiency is only poorly known, it was possible to
extract coherent cross section ratios for the different nuclear
excitation states. These ratios were used to scale the coherent
mutual-excitation cross section values from the minimum bias sample.
\Figref{run2_rho_y_dep} compares the obtained $y_\Pr$-dependence of
the coherent \Prz~production cross section to the three models.
Unfortunately the limited acceptance in $y_\Pr$ does not allow to
distinguish between different shapes of the distributions.

STAR has also measured the spin structure of the coherent
\Prz~production amplitudes in terms of spin density matrix
elements~\cite{run2_rho}. The matrix elements were extracted from the
minimum bias data by fitting the \Prz~decay angular distribution in
the \Prz~rest frame to the theoretical prediction for unpolarized
photoproduction~\cite{schilling_wolf}
\begin{multline}
  \frac{1}{\sigma}\, \frac{\D[2]{\sigma}}{\D{\cos\theta}\, \D{\phi}} =
  \frac{3}{4 \pi}\, \bigg[ \frac{1}{2} (1 - r_{00}^{04}) \\
  + \frac{1}{2} (3 r_{00}^{04} - 1) \cos^2\theta - \sqrt{2}\,
  \Re[r_{10}^{04}]\, \sin2\theta\, \cos\phi \\
  - r_{1\,-1}^{04}\,   \sin^2\theta\, \cos2\phi \bigg]
  \label{eq:rho_decay-distr}
\end{multline}
where~$\theta$ is the polar angle between the beam direction and the
direction of the~\Ppip\ and $\phi$~the azimuthal angle between the
decay and the production plane. Since STAR does not measure the
scattered beam particles, only an approximate production plane with
respect to the beam direction could be reconstructed. $r_{00}^{04}$
represents the probability that the~\Prz\ is produced with helicity~0,
$r_{1\,-1}^{04}$ the interference between helicity non-flip and
double-flip. Due to the ambiguity of photon source and target,
$\Re[r_{10}^{04}]$, which is related to the interference between
helicity non-flip and single-flip, could not be measured.
\Tabref{spin_matrix} summarizes the results. The measured spin density
elements are small, indicating $s$-channel helicity conservation. They
are also compatible with ZEUS measurements on \Prz~photoproduction from
proton targets~\cite{zeus_rho}.

\begin{table*}[htb]
  \renewcommand{\tabcolsep}{2pc}
  \renewcommand{\arraystretch}{1.2}
  \caption{Spin density matrix elements for coherent \Prz~production
    accompanied by mutual Coulomb excitation in \PAu-\PAu\ collisions
    at $\sqrtsnn = 200$~GeV~\cite{run2_rho} compared to ZEUS results
    from \Prz~production in $\PPh \Pp$
    collisions~\cite{zeus_rho}.}
  \begin{center}
    \begin{tabular}{rrr}
      \hline
      Parameter &
      \multicolumn{1}{c}{STAR} &
      \multicolumn{1}{c}{ZEUS} \\
      \hline
      
      $r_{00}^{04}$ &
      $-$0.03 $\pm$ 0.03$_\text{stat.} \pm$ 0.06$_\text{syst.}$ &
      0.01 $\pm$ 0.01$_\text{stat.} \pm$ 0.02$_\text{syst.}$ \\
      
      $\Re[r_{10}^{04}]$ &
      \multicolumn{1}{c}{---} &
      0.01 $\pm$ 0.01$_\text{stat.} \pm$ 0.01$_\text{syst.}$ \\
      
      $r_{1\,-1}^{04}$ &
      $-$0.01 $\pm$ 0.03$_\text{stat.} \pm$ 0.05$_\text{syst.}$ &
      $-$0.01 $\pm$ 0.01$_\text{stat.} \pm$ 0.01$_\text{syst.}$ \\
      \hline
    \end{tabular}
  \end{center}
  \label{tab:spin_matrix}
\end{table*}

\mathversion{bold}
\subsection{Interference Effects in Coherent \Prz~Production in \PAu-\PAu\ Collisions}
\label{subsec:rho_au_au_int}
\mathversion{normal}

In ultra-peripheral collisions the impact parameter~$b$ is typically
much larger than the sum of the nuclear radii of the beam particles.
The scattering of the virtual \Pqq\Pqqbar-pair on the target nucleus,
on the other hand, involves the short-ranged
($\mathcal{O}(1~\text{fm})$) strong interaction, so that the
production of the \Prz~mesons occurs in or very near to the target
nucleus. Furthermore photon source and target are indistinguishable,
so that either nucleus~1 emits a photon which produces a~\Prz\ at
nucleus~2, or vice versa. The system thus behaves like a two-source
interferometer with separation~$b$, where the interference between the
two possible amplitudes creates an entangled final state \Ppip\Ppim\
wave function. Since the~\Prz\ has negative parity, the two amplitudes
have different sign, so that the cross section can be written
as~\cite{rho_int}
\begin{equation}
  \sigma(b) = \lrabs{A(b, y_\Pr) - A(b, -y_\Pr)\, e^{i \vec{p}_T^{\,\Pr} \cdot \vec{b}}}^2
  \label{eq:rho_int}
\end{equation}
where $A(b, y_\Pr)$ and $A(b, -y_\Pr)$ are the \Prz~production
amplitudes for the two photon directions. The observed $\pT^\Pr$
spectrum is obtained by integrating \equref{rho_int} over~$\vec{b}$,
which is not measured in the experiment. For $y_\Pr = 0$ the two
amplitudes are equal so that
\begin{equation}
  \sigma(b) = \sigma_0(b) \lrBrk{1 - \cos(\vec{p}_T^{\,\Pr} \cdot \vec{b})}
  \label{eq:rho_int_zero_rap}
\end{equation}
with $\sigma_0(b)$ being the cross section without interference. From
\equref{rho_int_zero_rap} it is clear that the cross section is
suppressed for $\pT^\Pr \lesssim 2 \hbar / \left<b\right>$, where
$\left<b\right>$ is the mean impact parameter.

The $t$-spectra, with $t \approx \lrbrk{\pT^\Pr}^2$, are parametrized
by
\begin{equation}
  \frac{\D{N}}{\D{t}} = A\, e^{-B t} \lrBrk{1 + c (R(t) - 1)}
  \label{eq:rho_int_t}
\end{equation}
where $R(t) \equiv \lrBrk{\D{N} / \D{t}}_\text{int}^\text{MC} \Big/
\lrBrk{\D{N} / \D{t}}_\text{noint}^\text{MC}$ is the ratio of the
$t$-distributions with and without interference obtained from
Monte-Carlo simulations based on the KN~model~\cite{KN}. $A$~is the
overall normalization, $B$~the slope parameter for coherent
\Prz~production, and $c$~measures the strength of the interference: $c
= 0$ corresponds to no interference, whereas $c = 1$ is the expected
interference in the KN~model.

\begin{figure}[htb]
  \includegraphics[width=0.51\linewidth]{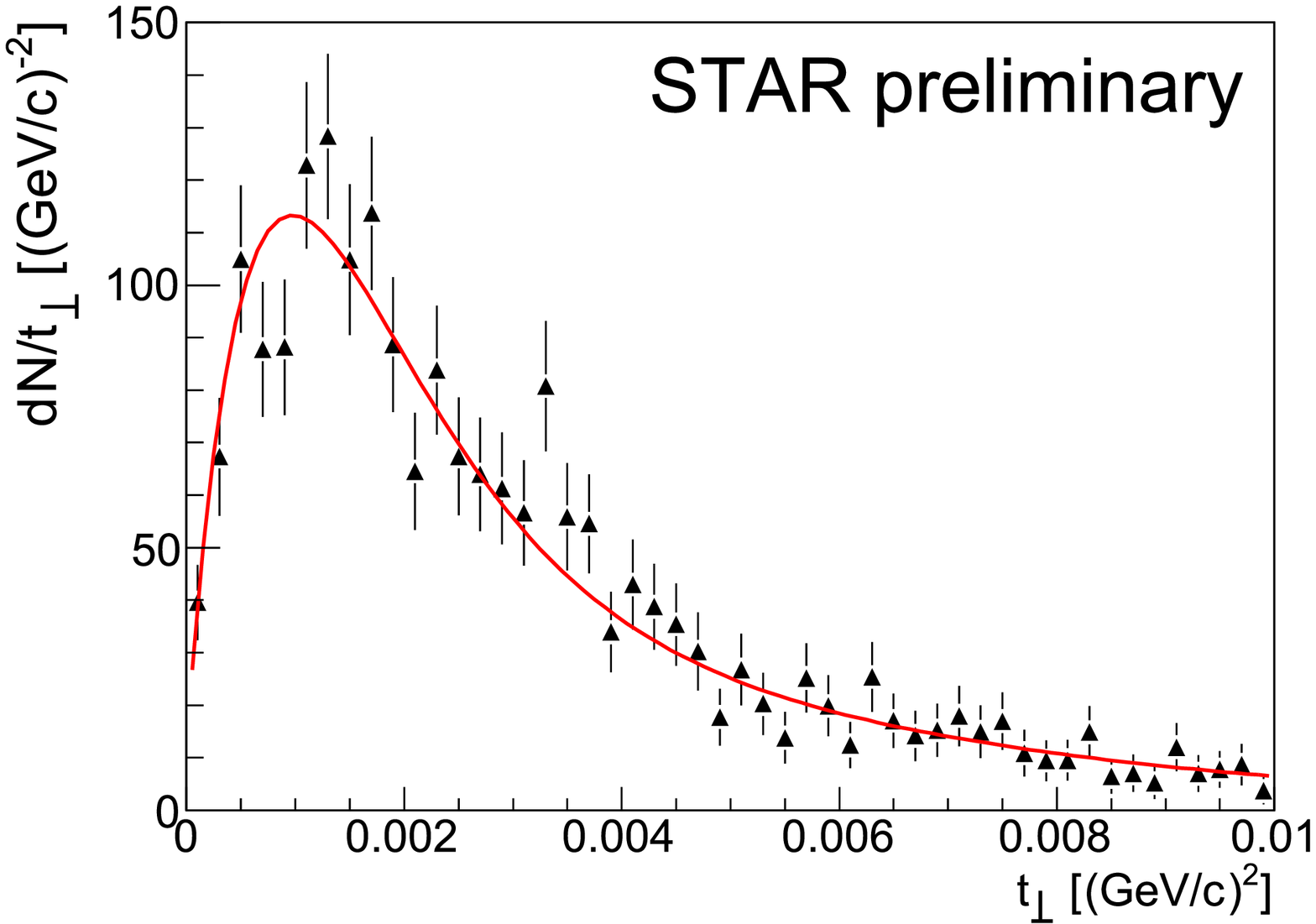}~\includegraphics[width=0.51\linewidth]{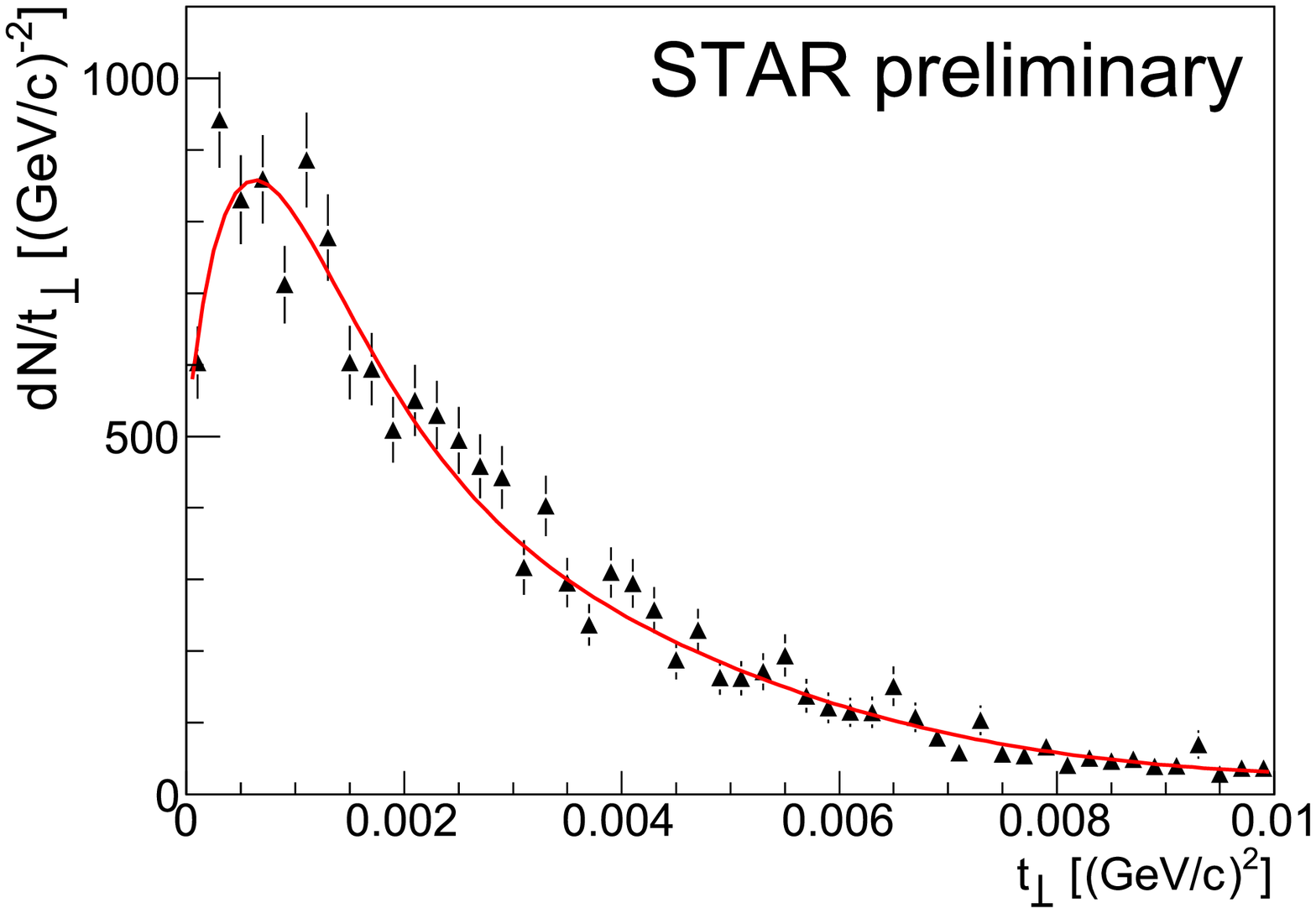}\\
  \centering a) \hspace{0.5\linewidth} b)
  \vspace{-3ex}
  \caption{Efficiency-corrected $t$-spectra for~(a) the minimum bias
    data set with $|y_\Pr| < 0.5$ and for~(b) the topology data set
    with $0.05 < |y_\Pr| < 0.5$. The minimum bias data exhibit a wider
    dip at low~$t$, because the median impact parameter is smaller.
    The solid line shows a fit to \equref{rho_int_t}.}
  \label{fig:run2_rho_interference}
\end{figure}

Due to the neutron tagging of mutual Coulomb excitation in the minimum
bias trigger, the median impact parameter of $\tilde{b} \approx 18$~fm
in this data set is smaller than the $\tilde{b} \approx 46$~fm in the
topology data. Therefore the dip in the minimum bias $t$-spectra
should extend to larger values of $t$. \Figref{run2_rho_interference}
shows the efficiency-corrected $t$-distributions for the run 2002
topology and minimum bias data sets from \PAu-\PAu\ collisions at
$\sqrtsnn = 200$~GeV in the central rapidity region $|y_\Pr| < 0.5$
together with a fit to \equref{rho_int_t}. To avoid contamination of
the topology sample by cosmic rays, in addition the region $|y_\Pr| <
0.05$ was excluded.

The minimum bias data exhibit an interference strength of $c = 0.92
\pm 0.07_\text{stat.}$ close to the expected value. The value from the
topology data set is lower with $c = 0.73 \pm 0.10_\text{stat.}$, but
could be affected by the imperfect topology trigger simulation.

\mathversion{bold}
\subsection{Incoherent \Prz~Production in \PAu-\PAu\ Collisions}
\label{subsec:rho_au_au_inc}
\mathversion{normal}

The incoherent \Prz~production, where the~\Prz\ couples to the
individual nucleons, can be measured by extending the analyzed range
of the \Prz~transverse momentum to 550\mevc. If the region of $t <
0.002\gevcsq$, where interference effects reduce the cross section
(see \subsecref{rho_au_au_int}), is excluded, the $t$-distribution can
be described by a double exponential function
\begin{equation}
  \frac{\D{\sigma}}{\D{t}} = A_\text{coh}\, e^{-B_\text{coh} t} + A_\text{inc}\, e^{-B_\text{inc} t}
  \label{eq:rho_t_dep}
\end{equation}

$B_\text{coh}$ is the slope parameter of the exponential that
dominates the low-$\pT^\Pr$ region and corresponds to the coherent
production. Accordingly $B_\text{inc}$ is the slope parameter of the
incoherent production that extends to larger~$\pT^\Pr$.

\begin{figure}[htb]
  \begin{center}
    \includegraphics[width=0.8\linewidth]{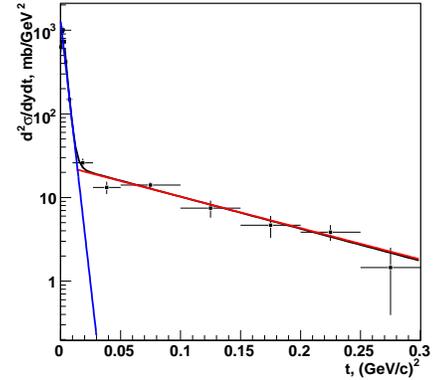}
  \end{center}
  \vspace{-5ex}
  \caption{Dependence of the \Prz~production cross section on the
    squared momentum transfer~$t$ for the minimum bias
    data~\cite{run2_rho}. The solid line shows the result of a fit to
    the double exponential \equref{rho_t_dep} with the steep line at
    low~$t$ corresponding to coherent production. The line at
    larger~$t$ describes the incoherent \Prz~production.}
  \label{fig:run2_rho_t_spectrum}
\end{figure}

\Figref{run2_rho_t_spectrum} shows a fit of \equref{rho_t_dep} to the
year 2002 minimum bias data from \PAu-\PAu\ at $\sqrtsnn = 200$~GeV
that gives $B_\text{coh} = 388 \pm 24_\text{stat.}~(\text{GeV}\! /\!
c)^{-2}$ and $B_\text{inc} = 8.8 \pm 1.0_\text{stat.}~(\text{GeV}\!
/\! c)^{-2}$~\cite{run2_rho}. By integrating the exponentials the
cross section ratio of incoherent to coherent \Prz~production
accompanied by mutual Coulomb excitation was measured to
\measresult{29}{3}{8}{\%}.

\mathversion{bold}
\subsection{\Prz~Production in \Pd-\PAu\ Collisions}
\label{subsec:rho_d_au}
\mathversion{normal}

In \Pd-\PAu\ ultra-peripheral collisions the virtual photon is
predominantly emitted by the gold nucleus, due to its larger charge.
There are two possible \Prz~production processes: coherent scattering
on the entire deuteron, $\PvPh \Pd \to \Prz \Pd$, and ``incoherent''
scattering on the individual nucleons of the deuteron in the course of
which the deuteron dissociates, $\PvPh \Pd \to \Prz\, \Pp\Pn$.

During the year 2003 run STAR took data of \Pd-\PAu\ collisions at
$\sqrtsnn = 200$~GeV using two UPC triggers. The first trigger was a
topology trigger which selected mainly coherent interactions. The
second trigger required, in addition to the topology condition, a
neutron signal from the deuteron breakup in the corresponding ZDC.
This data sample contains incoherent scattering events.

\begin{figure}[htb]
  \begin{center}
    \includegraphics[width=0.8\linewidth]{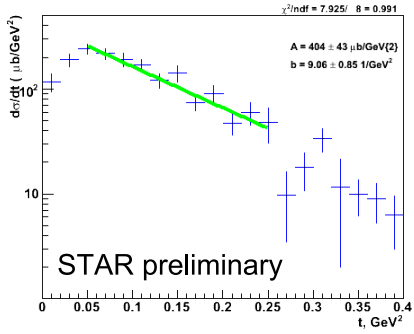}
  \end{center}
  \vspace{-5ex}
  \caption{$t$-distribution for incoherent \Prz~production accompanied
    by deuteron dissociation. The solid line shows an exponential
    fit.}
  \label{fig:dau_rho_t_incoh}
\end{figure}

\Figref{dau_rho_t_incoh} shows the $t$-spectrum, with $t \approx
\lrbrk{\pT^\Pr}^2$, for incoherent scattering accompanied by deuteron
dissociation. A fit with an exponential gives a slope parameter of $B
= 9.1 \pm 0.9_\text{stat.}~(\text{GeV}\! /\! c)^{-2}$ which is related
to the nucleon form factor. This value is compatible with the slope
parameter of $B_\text{inc} = 8.8 \pm 1.0_\text{stat.}~(\text{GeV}\!
/\! c)^{-2}$ measured in incoherent \Prz~production accompanied by
mutual Coulomb excitation in \PAu-\PAu\ collisions at $\sqrtsnn =
200$~GeV (cf. \subsecref{rho_au_au_inc};~\cite{run2_rho}) and also
agrees with the value of $B = 10.9 \pm 0.3_\text{stat.} {+1.0 \atop
  -0.5}\!\, _\text{syst.}~(\text{GeV}\! /\! c)^{-2}$ measured at
ZEUS~\cite{zeus_rho}. The downturn of the distribution at low~$t$,
which is also seen in low energy \PPh\Pd\ experiments~\cite{slac}, is
due to the fact that there is not enough energy to dissociate the
deuteron.

\mathversion{bold}
\section{\Pep\Pem-Pair Production in Photon-Photon Interactions}
\label{sec:ee_production}
\mathversion{normal}

During the year 2002 run STAR also measured the production of
\Pep\Pem-pairs in photon-photon interactions in ultra-peripheral
\PAu-\PAu\ collisions at $\sqrtsnn = 200$~GeV~\cite{star_ee}. Since
the electromagnetic field is strong with $Z \alpha \approx 0.6$, one
expects to see higher-order QED effects. In order to enrich collisions
at small impact parameters, where the fields are strongest and thus
higher-order corrections should be most pronounced, a ZDC trigger was
employed, requiring mutual nuclear dissociation and thus limiting the
impact parameter to $2 R_\PA < b \lesssim 30$~fm.

In the offline analysis events were selected that had two
reconstructed tracks of opposite charge emerging from a common
(primary) vertex along the beamline. Up to two additional tracks, not
associated to the primary vertex, were allowed in order to account for
random backgrounds. Most of the~\Pepm\ are produced in forward
direction with low transverse momentum, so that only a small fraction
of the pairs reached the central detectors, even though the solenoid
field was reduced to 0.25~T, half of its nominal value. The tracks
were required to have pseudorapidities of $|\eta| < 1.15$ and
transverse momenta of $65 < \pT < 130\mevc$. The latter cut ensured
that the \Pepm~identification via the specific energy loss in the TPC
gas had only minimal contaminations from other particle species.
Additional cuts on the pair transverse momentum, $\pT^{\Pep\!\Pem} <
100\mevc$, and the pair rapidity, $|y_{\Pep\!\Pem}| < 1.15$, gave a
final sample of 52~events in the pair invariant mass region of $140 <
M_{\Pep\!\Pem} < 265\mevcc$.

\begin{figure}[htb]
  \includegraphics[width=0.495\linewidth]{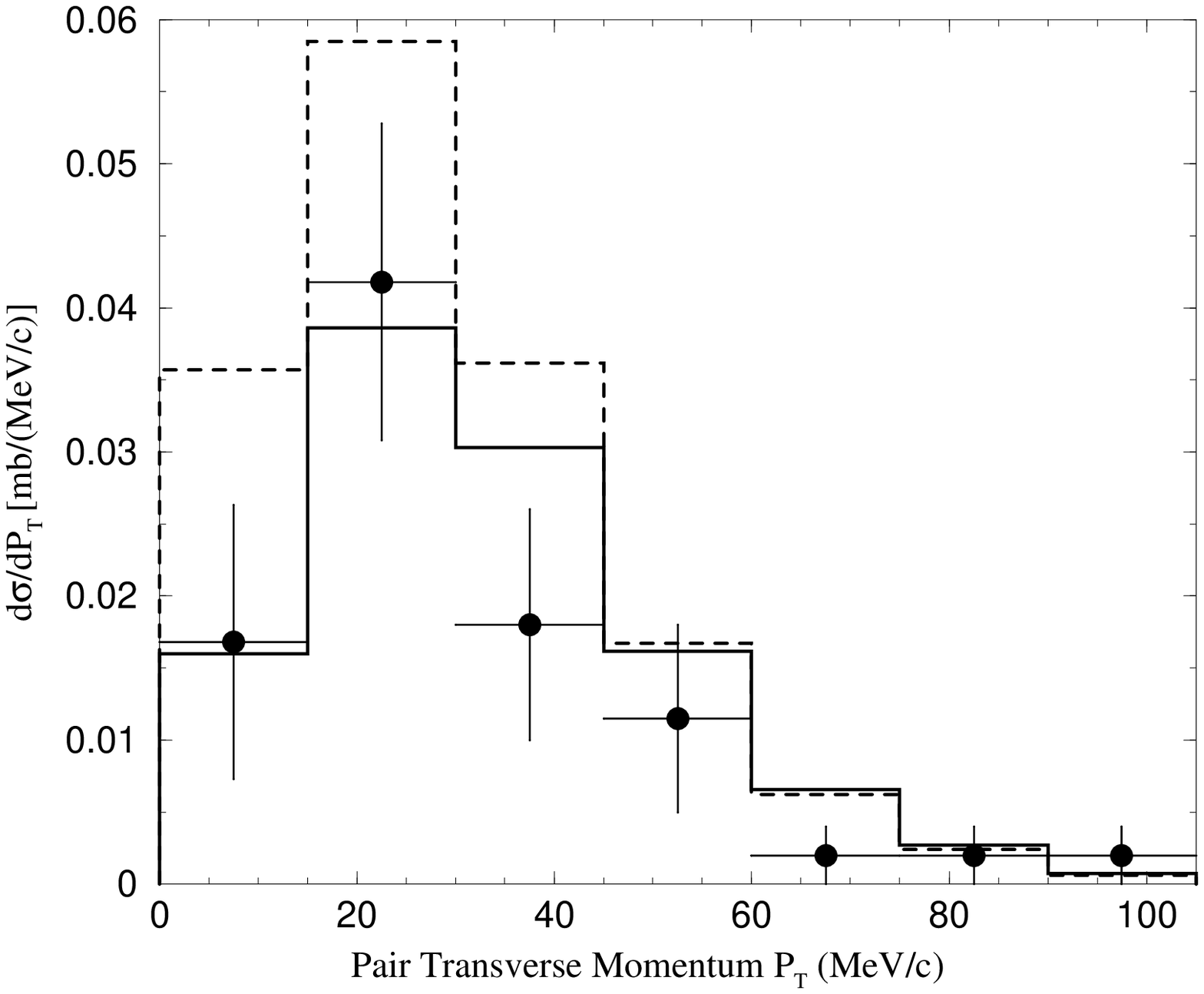}~\includegraphics[width=0.495\linewidth]{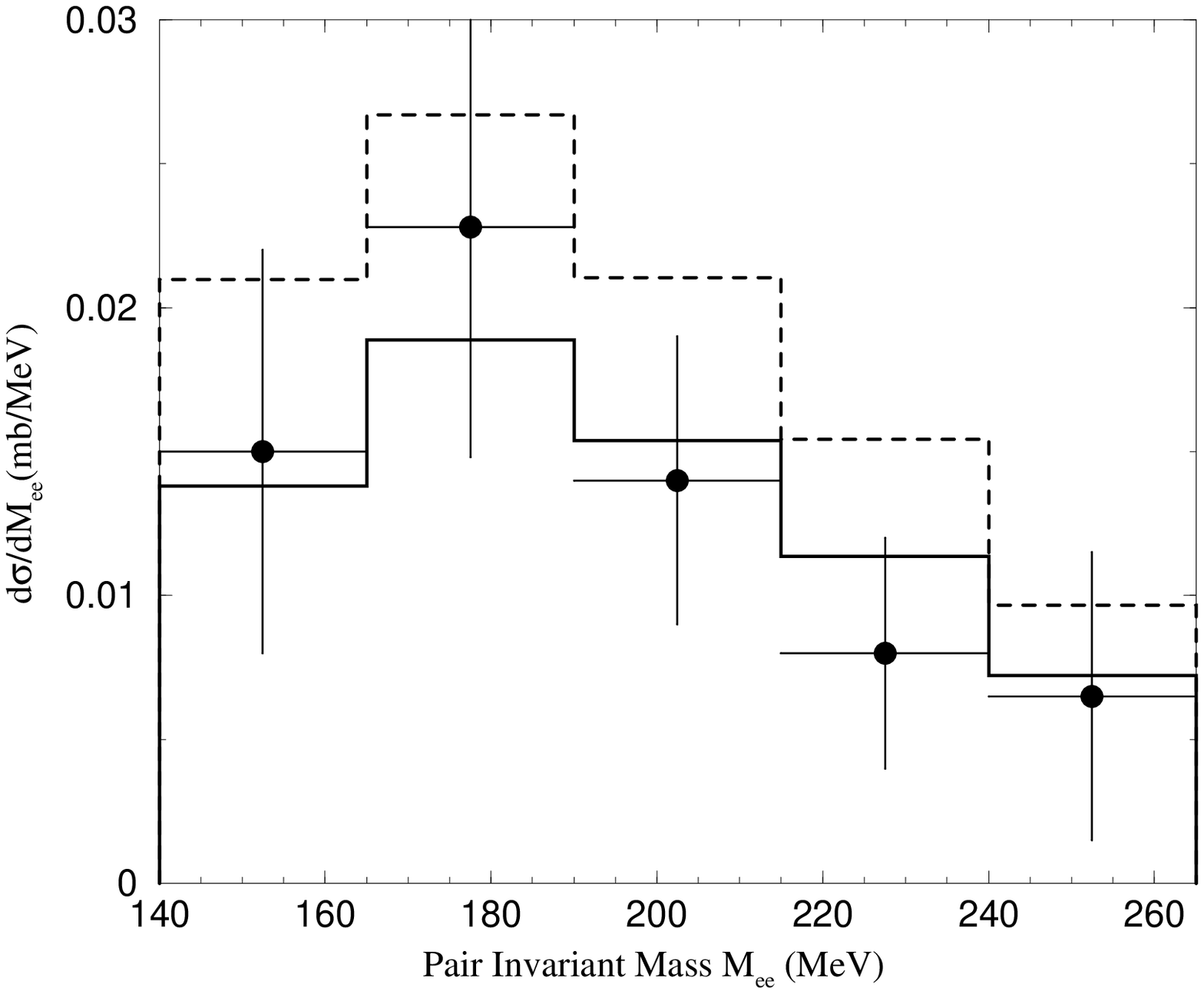}\\
  \centering a) \hspace{0.5\linewidth} b)
  \vspace{-3ex}
  \caption{(a)~Pair transverse momentum spectrum. (b)~\Pep\Pem\
    invariant mass distribution. The dashed line shows the lowest
    order, the solid line the higher-order QED calculation according
    to~\cite{baltz_ee}.}
  \label{fig:ee_pair_baltz}
\end{figure}

The measured cross section in the selected kinematic range is
\measresult{1.6}{0.2}{0.3}{mb}. \Figref{ee_pair_baltz}a shows the
cross section as a function of the transverse momentum of the
\Pep\Pem-pair, \figref{ee_pair_baltz}b the pair invariant mass
distribution. The data are compared to a a recent QED calculation of
Baltz that includes a realistic phenomenological treatment of the
nuclear Coulomb excitation~\cite{baltz_ee}. The lowest order QED
(dashed line in \figref{ee_pair_baltz}) overpredicts the data. The
calculated cross section of 2.34~mb is nearly 2~standard deviations
above the data. The corresponding higher-order QED calculation gives a
cross cestion value of 1.67~mb, compatible with the measurement, and
also describes the kinematic distributions well.

\mathversion{bold}
\section{Coherent Photonuclear \Ppip\Ppim\Ppip\Ppim\ Production}
\label{sec:4pi_production}
\mathversion{normal}

During the year 2004 run STAR took data in \PAu-\PAu\ collisions at
$\sqrtsnn = 200$~GeV using a ``multi-prong'' trigger which combined
the UPC neutron tag in form of coincident ZDC signals with a low hit
multiplicity in the CTB. In addition a veto from the large-tile
Beam-Beam Counters (BBCs)~\cite{star_bbc}, which cover $2.1 < |\eta| <
3.6$, was applied.

Events were selected that have four charged tracks coming from one
common (primary) vertex. These four-prongs were required to be neutral
and to have a net transverse momentum below 150\mevc, thus selecting
coherent production. A low number of additional tracks, unassociated
to the primary vertex, were allowed in the events to account for
backgrounds.

\begin{figure}[htb]
  \includegraphics[width=0.52\linewidth]{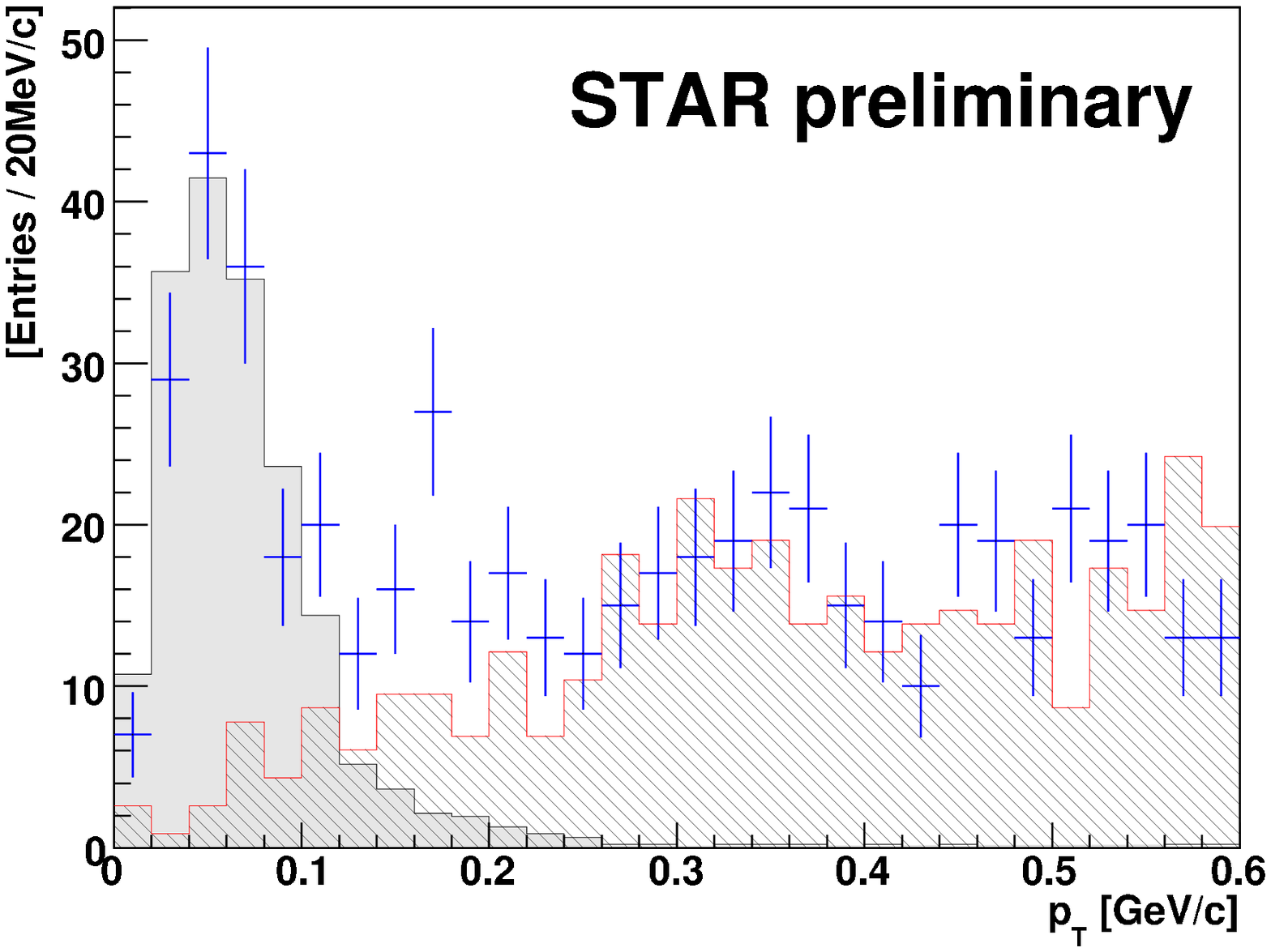}~\includegraphics[width=0.52\linewidth]{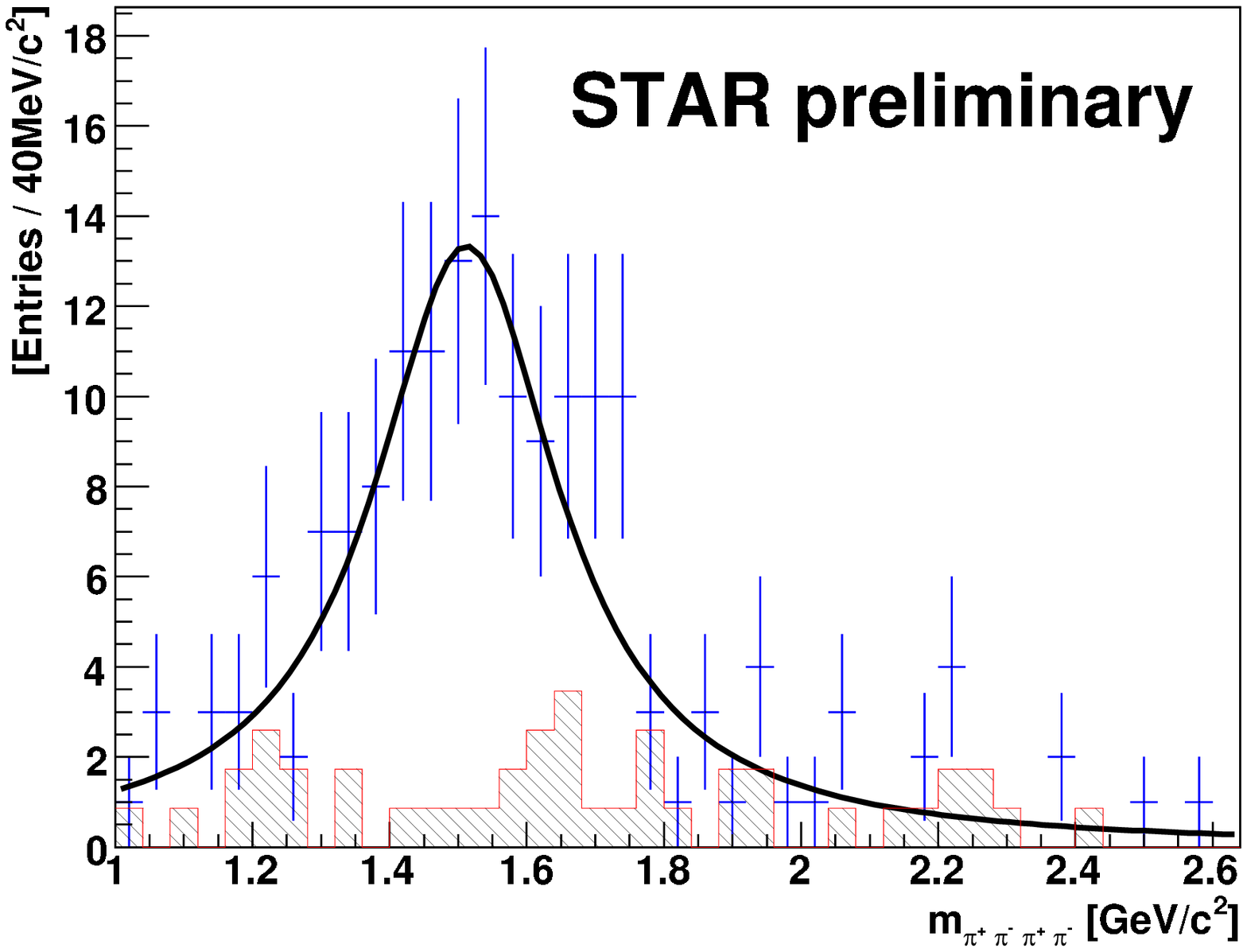}\\
  \centering a) \hspace{0.5\linewidth} b)
  \vspace{-3ex}
  \caption{(a)~Transverse momentum spectrum of coherently produced
    neutral four-prongs. The points are the data, the shaded histogram
    depicts the expected distribution from simulation. The hatched
    histogram shows the background estimated from charged four-prongs.
    (b)~\Ppip\Ppim\Ppip\Ppim\ invariant mass distribution of
    coherently produced four-prongs. A broad peak indicates resonant
    \Ppip\Ppim\Ppip\Ppim\ production.}
  \label{fig:run4_rhoprime}
\end{figure}

The transverse momentum distribution of the neutral four-prongs in
\figref{run4_rhoprime}a shows the expected peak at low~\pT,
characteristic for coherent production. The \Ppip\Ppim\Ppip\Ppim\
invariant mass spectrum in \figref{run4_rhoprime}b shows a broad peak
of approximately 100~events around 1510\mevcc\ which could be caused
by the production of excited \Prz\ states like the $\Prz(1450)$ and/or
the $\Prz(1700)$.

\end{document}